\documentclass[3p]{elsarticle}
\usepackage{graphicx}
\usepackage{epstopdf}
\usepackage{epsfig}
\usepackage{mathrsfs}
\usepackage{amssymb}
\usepackage{upgreek}
\usepackage{bbold}
\usepackage{color}
\usepackage{bm}
\usepackage{longtable}
\usepackage{orcidlink}

\begin{document}

\title{Coherence thermometry using multipartite quantum systems}

\author[1]{Pranav Perumalsamy}
\ead{pranavpp747@gmail.com}
\affiliation[1]{organization={Department of Artificial Intelligence and Data Sciences}, addressline={Chennai Institute of Technology, Chennai 600069}, country={India}}

\author[2]{Abhijit Mandal \orcidlink{0000-0001-7101-9495}}
\ead{a.mandal1.tmsl@ticollege.org}
\affiliation[2]{organization={Department of Mathematics, Techno Main Salt Lake}, addressline={Techno India Group, EM 4/1, Sector V, Salt Lake, Kolkata  700091}, country={India}}

\author[3]{Sovik Roy \corref{cor1} \orcidlink{0000-0003-4334-341X}}
\ead{s.roy2.tmsl@ticollege.org}
\affiliation[3]{organization={Department of Mathematics, Techno Main Salt Lake}, addressline={Techno India Group, EM 4/1, Sector V, Salt Lake, Kolkata  700091}, country={India}}
\cortext[cor1]{Corresponding author}

\author[4]{Md Manirul Ali \orcidlink{0000-0002-5076-7619}}
\ead{manirul@citchennai.net}
\affiliation[4]{organization={Centre for Quantum Science and Technology}, addressline={Chennai Institute of Technology, Chennai 600069}, country={India}}

\date{\today}

\begin{abstract}
\noindent Accurate temperature measurement at the quantum scale is becoming increasingly important for emerging 
quantum technologies, motivating the development of quantum thermometry based on quantum resources. In this work, 
we investigate how finite environmental temperature influences the coherence dynamics of multipartite quantum systems 
and examine whether quantum coherence can serve as a temperature sensitive observable. We consider a tripartite 
spin-boson model interacting with finite temperature dephasing environments under two physically 
distinct reservoir configurations, namely local and common environments. The dynamics of representative tripartite 
pure and mixed states are quantified using the relative entropy of coherence. Our results show that local dephasing 
produces a universal monotonic decay of coherence, with increasing temperature accelerating decoherence for all states. 
In contrast, common dephasing generates a markedly state dependent thermal response. Under common dephasing, 
the $\vert GHZ \rangle$ and $\vert Star\rangle$ states undergo complete coherence loss, the $\vert W\rangle$ state 
exhibits temperature independent stationary coherence, and the $\vert W\overline{W}\rangle$ state retains finite 
residual coherence at long times. Similar state dependent behaviour is also observed for mixed states. These results 
demonstrate that the thermal susceptibility of quantum coherence is governed jointly by the environmental 
configuration and the internal architecture of the multipartite quantum state. Furthermore, we establish a direct 
coherence temperature correspondence through representative thermometry tables, providing a \textit{proof-of-principle} 
foundation for coherence based quantum thermometry.
\end{abstract}



\maketitle

\section{Introduction}\label{sec:intro}

\noindent Quantum coherence is one of the most fundamental manifestations of quantum mechanics and constitutes an indispensable
resource for modern quantum technologies, including quantum computation, quantum communication, quantum sensing, and
quantum metrology \cite{nielsen2010quantum,baumgratz2014quantifying,streltsov2017colloquium,horodecki2009quantum,
degen2017quantum,giovannetti2011advances,rahman2025gravitational}. The coherent superposition of quantum states
enables the generation of entanglement and other nonclassical correlations that provide significant advantages over their
classical counterparts in a wide variety of quantum information processing tasks. The successful realization of scalable
quantum technologies therefore relies critically on preserving quantum coherence over sufficiently long time scales. In realistic
physical systems, however, quantum systems are inevitably coupled to surrounding environments. Such unavoidable
system--environment interactions lead to decoherence, progressively destroying quantum coherence and thereby limiting
the performance and reliability of quantum devices
\cite{zurek2003decoherence,breuer2002theory,weiss2012quantum,breuer2016,devega2017,singh2024comprehensive,
bachain2026quantum,abd2019restraining,abd2022quantum}. Decoherence also degrades the quality
of quantum channels and reduces the fidelity of quantum information protocols such as quantum teleportation and quantum cryptography \cite{bennett1993teleporting,horodecki1996teleportation,horodecki1999general,gisin2002quantum,yu2004finite}.
Consequently, understanding the mechanisms responsible for coherence degradation and developing effective strategies
for protecting quantum coherence remain central challenges in quantum information science.\\

\noindent Since realistic quantum systems are inherently open, their dynamics generally exhibit nonunitary evolution arising 
from interactions with their surrounding environments. Understanding and controlling decoherence in such systems 
has therefore been a central topic in the theory of open quantum systems \cite{breuer2002theory,weiss2012quantum}. 
While conventional treatments often assume that environmental correlations decay rapidly compared with the characteristic 
timescales of the system, a wide range of contemporary quantum platforms operate beyond this approximation, exhibiting 
dynamics that depend explicitly on the history of their interactions with the environment \cite{breuer2016,de2017dynamics}. 
Such memory effects are particularly pronounced in structured environments, where environmental correlations can persist 
over timescales comparable to those of the system. They have been investigated and observed in a variety of physical platforms, 
including photonic systems, superconducting circuits, solid-state spin qubits, hybrid quantum architectures, and quantum 
transport devices \cite{bellomo2007non,wang2008decoherence,dajka2008non,paz2008dynamics,li2010entanglement}. 
Beyond their role in modifying decoherence dynamics, structured environments provide valuable opportunities for reservoir 
engineering and coherent control. By tailoring the environmental spectral density and exploiting reservoir memory 
effects, it is possible to mitigate decoherence and enhance the preservation of quantum coherence and other quantum resources \cite{braun2002creation,mazzola2009sudden,aliPRA2010decoherence,sarlette2011stabilization,
liu2011experimental,laine2012nonlocal,bylicka2014non,brito2015knob,nokkala2016complex,jaloum2025work}. \\

\noindent Among the various environmental parameters governing open-system dynamics, temperature plays a
particularly important role because it directly determines the strength of thermal fluctuations experienced by the
quantum system. Accurate temperature measurement at the nanoscale has consequently become an important
objective in modern quantum technologies, stimulating significant advances in quantum thermometry
\cite{stace2010quantum,jevtic2015single,correa2015individual,xiao2024construction}. Unlike conventional thermometers,
quantum thermometric schemes exploit uniquely quantum resources, such as quantum coherence, entanglement,
and quantum correlations, to estimate temperature with high sensitivity in regimes where classical approaches
become inadequate. Such techniques provide powerful tools for probing thermal fluctuations, energy transport,
and microscopic heat exchange in nanoscale quantum devices.\\

\noindent From the perspective of open quantum systems, environmental temperature is not merely a source of
decoherence but also a physically meaningful control parameter that directly influences the system dynamics. In
finite temperature dephasing models, the time dependent  dephasing rates appearing in the quantum master
equation depend explicitly on thermal factors associated with the reservoir. Consequently, the evolution of
quantum coherence becomes intrinsically linked to the surrounding temperature, establishing a direct
correspondence between coherence dynamics and the thermal environment. This observation naturally raises
the possibility of employing quantum coherence itself as a temperature sensitive observable. If the coherence
is measured at a fixed evolution time, the environmental temperature can, in principle, be inferred from the
measured coherence. Such a coherence temperature correspondence provides the physical basis for
coherence based quantum thermometry and is closely related to emerging concepts in quantum calorimetry,
where engineered quantum systems serve as microscopic probes of energy exchange and heat fluctuations.\\

\noindent Although the dynamics of quantum coherence under various decoherence mechanisms have been
extensively investigated, most previous studies have focused primarily on the influence of environmental
structure, \textit{system-environment} coupling, or coherence preservation. Comparatively little attention has
been devoted to systematically understanding how finite environmental temperature modifies multipartite
quantum coherence in different reservoir configurations. In multipartite systems, the response of coherence
to thermal fluctuations is expected to depend not only on the reservoir properties but also on the internal
architecture and symmetry of the underlying quantum state. Identifying such state dependent thermal susceptibility
is important both for understanding decoherence in realistic quantum devices and for assessing the suitability
of different multipartite states as temperature sensitive quantum probes.\\

\noindent Motivated by these considerations, we investigate the temperature-dependent coherence dynamics
of representative multipartite quantum states evolving under local and common finite temperature time-dependent
dephasing environments. We consider four representative tripartite pure states, namely the $\vert GHZ\rangle$,
$\vert W\rangle$, $\vert W\overline{W}\rangle$, and $\vert Star \rangle$ states, together with physically relevant
mixed states constructed from them. The coherence dynamics are quantified using the relative entropy of coherence.
By systematically comparing local and common dephasing environments over a broad temperature range, we
demonstrate that the thermal susceptibility of multipartite coherence is governed jointly by the environmental
configuration and the internal architecture of the quantum state. Furthermore, we establish a direct
coherence temperature correspondence through representative thermometry tables, thereby providing a
\textit{proof-of-principle} demonstration that quantum coherence can serve as a temperature sensitive observable in
finite temperature environments. Our results provide new insight into thermal decoherence in
multipartite quantum systems and identify coherence as a promising resource for future coherence based
quantum thermometry and nanoscale quantum calorimetry.\\

\noindent The paper is organized as follows. In Sec.~2 we briefly review the relative entropy of coherence used
throughout this work. Section~3 introduces the tripartite spin-boson models under local and common finite temperature
dephasing environments. Section~4 presents the multipartite states considered in the analysis together with
representative coherence thermometry tables. The temperature-dependent coherence dynamics of both pure and
mixed states are discussed in Sec.~5. Finally, Sec.~6 summarizes the principal results and outlines possible future
research directions.

\section{Measure of coherence:}
\noindent Quantum coherence is a fundamental resource for emerging quantum technologies, and its quantification has only recently been formalized within a unified resource-theoretic framework \cite{baumgratz2014quantifying,streltsov2017colloquium}. This resulted in several fundamental advancements in the field of resource theory of quantum coherence \cite{baumgratz2014quantifying, winter2016, radha2016, chitambar2019}. To quantify coherence of a state we consider the well-known measure of relative entropy of coherence. We consider here the coherence with respect to computational basis of the multiqubit system.
For any quantum state $\rho$, the relative entropy of coherence is given by \cite{baumgratz2014quantifying}
\begin{eqnarray}
\label{coher1}
C_R(\rho) &= \min\limits_{\sigma \in \mathcal{I}} S(\rho \| \sigma),
\end{eqnarray}
where $\mathcal{I}$ denote the set of incoherent states. The relative entropy $S(\rho \| \sigma)
=\rm{Tr} \left(\rho \ln \rho-\rho \ln \sigma \right)$. For the relative entropy of coherence, it has been
shown that the closest incoherent state is simply the diagonal part of the density matrix, denoted by
$\rho_{diag}$, obtained by removing all off-diagonal elements of $\rho$ in the reference basis.
Hence it is not necessary to perform the minimization to determine the quantum coherence. The
relative entropy of coherence Eq.(\ref{coher1}) reduces to the form
\begin{eqnarray}
\label{coher2}
C_R(\rho) &= S(\rho_{diag}) - S(\rho),
\end{eqnarray}
where $S(\rho)$ is the von Neumann entropy of the state $\rho$. When a coherent quantum system interacts with surrounding environment, usually it suffers from decoherence process \cite{breuer2002theory}. Recently, the dynamics of quantum coherence have been explored in open quantum systems \cite{radha20192363,cao2020fragility,ali2022mm,ali2023mm}. Our results suggest that the temperature-dependent dynamics of quantum coherence can be utilized for quantum thermometry, where the temperature of the surrounding environment is inferred from the evolution of coherence.

\section{Description of the models}

\begin{figure*}[h]
\includegraphics[width=16.16cm]{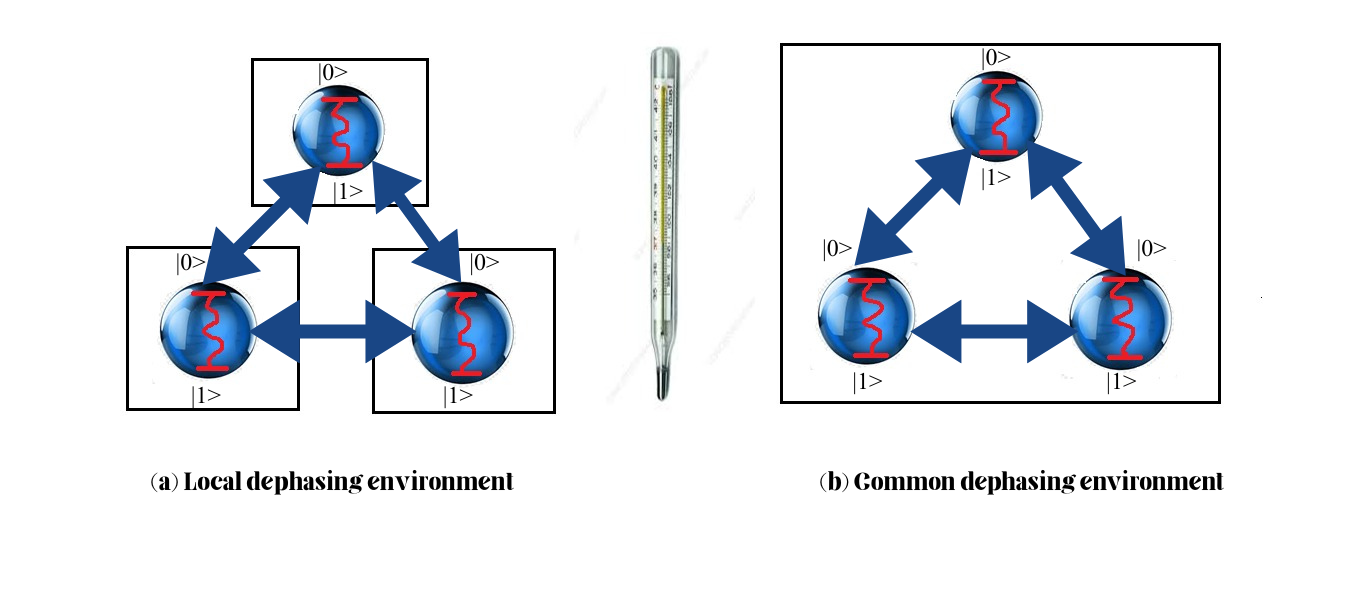}
\caption{Schematic illustration of a tripartite qubit system subjected to (a) local dephasing and (b) common dephasing environments. In the local configuration, each qubit interacts independently with its own bosonic bath (represented by three squares in fig.~(a)) while the spheres are representing the qubits, leading to uncorrelated dephasing. In contrast, in the common environment, all qubits are collectively coupled to a single bosonic reservoir (single square in fig.~(b)), giving rise to correlated dephasing dynamics. In both cases, the environments are assumed to be at finite temperature. }
\label{fig1}
\end{figure*}

\noindent We discuss the three-qubit models subjected to two types of noisy environments, namely (i) Local Dephasing Environment (LDE) and (ii) Common Dephasing Environment (CDE), as illustrated in Fig. \ref{fig1}. In the local dephasing model, each qubit interacts with its own independent bosonic reservoir. Consequently, the environmental fluctuations acting on different qubits are statistically independent and uncorrelated. This configuration models situations in which the qubits are spatially separated or coupled to distinct environments, such that each qubit undergoes independent phase randomization without exchanging information through the environment. Local dephasing therefore represents the standard scenario of independent decoherence widely studied in open quantum systems \cite{dajka2008non,romero2008prl,wein2009stein,zoushao2010,radha20192363}.
In contrast, the common dephasing model describes the situation in which all qubits interact collectively with the same bosonic reservoir. In this case, the environmental fluctuations are shared by all qubits, generating correlated phase noise and collective decoherence effects. Such correlated \textit{system-environment} interactions can give rise to decoherence-free subspaces, stationary coherence, and other collective phenomena that are absent in independent reservoirs \cite{paz2008dynamics,mazzola2009sudden,aliPRA2010decoherence,xzhao2013,wangji2015}. The theoretical model considered in this
work can be realized in several state-of-the-art quantum platforms, including superconducting qubits \cite{berger2015,burnett2019,odeh2025}, trapped-ion systems \cite{kiel2001,bara2011,sun2025}, Nitrogen vacancy centers in
diamond \cite{cory2010,yichou2015}, and photonic quantum-information processors \cite{bh2011liu,cao2020fragility}. In these systems, multipartite entangled states such as $\vert GHZ \rangle$ and $\vert W \rangle$
states can be prepared with high fidelity, while local and collective dephasing environments can be realized using reservoir engineering,
or structured bosonic environments. In addition to dedicated experimental platforms, modern quantum computing and quantum simulation architectures provide versatile testbeds for investigating open quantum systems \cite{garciaperez2020}.

\subsection{Three Qubits Under Local Dephasing Environment}\label{subsec:Local}

\noindent We consider a spin-boson framework consisting of three non-interacting qubits, each coupled to its own local bosonic environment, as shown in Fig.~\ref{fig1} . The total Hamiltonian for the three qubits and their environment is
\begin{eqnarray}
\label{totalhamiltonianldp}
H = \sum_{i=1}^3\Big[\frac{\hbar}{2}\omega_{0}^i\sigma_{z}^i + \sum_{k}\hbar \omega_{ik}b_{ik}^{\dagger}b_{ik} + \sigma_{z}^i
(B_{i} + B_{i}^{\dagger})\Big],
\end{eqnarray}
where $B_{i}=\hbar \sum_{k} g_{ik}b_{ik}$. In this model, $\sigma_{z}^i$ and $\omega_{0}^i$ denote the Pauli spin operator and the transition frequency of the $i^{th}$ qubit, respectively. For simplicity, all three qubits are taken to have the same transition frequency $\omega_0$. Each local environment associated with the $i^{th}$ qubit is modeled as a collection of bosonic modes with frequencies $\omega_{ik}$; the operators $b_{ik}^{\dagger}$ and $b_{ik}$ are the creation and annihilation operators associated to the $k^{th}$
mode of this local environment interacting with the $i^{th}$ qubit, and $g_{ik}$ characterizes the coupling strength between the $i^{th}$
qubit and the $k^{th}$ mode of that local environment. Initially, the three-qubit system is assumed to be decoupled from its environment
and each local environment is initially in thermal equilibrium at a temperature $T_i$. The subsequent dynamics generated by the total Hamiltonian $H$ are obtained by tracing out the environmental degrees of freedom, yielding the reduced density matrix of the three-qubit system. For the tripartite system, the quantum master equation describing the decay dynamics under local dephasing is given by
\begin{eqnarray}
\label{NL}
\frac{d}{dt} \rho(t) = -\frac{i}{\hbar} \big[ H_S, \rho(t) \big] +
\sum_{i=1}^{3} \gamma_{i}(t)  \Big( \sigma_z^i \rho(t) \sigma_z^i - \rho(t) \Big),
\end{eqnarray}
where the system Hamiltonian is
\begin{eqnarray}
\label{Hs}
H_S=\frac{\hbar}{2} \sum_{i=1}^3 \omega_0 \sigma_{z}^i
\end{eqnarray}
and the time dependent  dephasing rate is given by
\begin{eqnarray}
\label{gammat}
\gamma_{i}(t) = 2 \int_{0}^{\infty} d\omega  J_{i}(\omega) \coth\left(\frac{\hbar \omega}{2 k T_i}\right) \frac{\sin(\omega t)}{\omega}.
\end{eqnarray}
Here $J_i(\omega)=\sum_k |g_{ik}|^2 \delta(\omega-\omega_{ik})$ is the spectral density of the local environment for the
$i^{th}$ qubit. In the continuum limit, the discrete mode index is replaced by a continuous frequency variable, such that
$g_{ik} \rightarrow g_i(\omega)$ and the summation transforms into an integral weighted by the density of states.
As a result, the spectral density can be expressed as $J_i(\omega)=P_i(\omega)\vert g_i(\omega) \vert^2$, where
$P_i(\omega)$ is the density of states of the $i^{th}$ reservoir. The time dependent  dephasing rate $\gamma_{i}(t)$ is
fully determined by the spectral density $J_{i}(\omega)$. For the current model, we consider an Ohmic spectral density \cite{leggett1987dynamics} given by
\begin{eqnarray}
J_{i}(\omega) = \eta_{i} \omega \exp\left(-\frac{\omega}{\Lambda_i} \right),
\label{ohm}
\end{eqnarray}
where $\eta_i$ represents system-reservoir coupling strength and $\Lambda_i$ is the cutoff frequency. Typically, the
environment is considered to be large enough to rapidly returns to its initial state. 
Here, the influence of the structured spectral density $J_{i}(\omega)$ on the coherence dynamics of the three-qubit 
system is explicitly taken into account. The resulting time-dependent dephasing rate reflects the finite correlation 
properties of the reservoir and depends explicitly on the chosen spectral density and temperature. Usually, quantum master 
equations are derived under short environment correlation time \cite{scully1997quantum,carmichael2013statistical,gardiner2004quantum}, 
characterized by a bath relaxation time that is significantly shorter than the system's evolution time. 
In this approximated regime, memory effects in the environment are negligible. However, if the bath relaxation time 
becomes comparable to (or longer than) the system timescale, environmental
memory effects can no longer be ignored. In such cases, the decoherence dynamics of the three-qubit system must be
described using the quantum master equation in Eq.~(\ref{NL}), which incorporates time dependent  decay
rates $\gamma_{i}(t)$ as given in Eq.~(\ref{gammat}). \\

\subsection{Three Qubits Under Common Dephasing Environment}\label{subsec:Common}

\noindent Now, we consider the model where a spin-boson framework consisting of all three qubits is under the interaction with a common reservoir as shown in Fig.~\ref{fig1} .  The total Hamiltonian for the three qubits coupled to a common environment is
\begin{eqnarray}
\label{Dephc}
H = \frac{\hbar}{2} \sum_{i=1}^{3} \omega_0^i \sigma_z^i + \sum_k \hbar \omega_{k} b_{k}^{\dagger} b_{k}
+ \hbar S_z \sum_k \left( g_k b_{k}^{\dagger} + g_k^{\ast} b_{k} \right),
\end{eqnarray}
where $S_z = \sum_i \sigma_z^i$ denotes the collective spin operator for the tripartite qubit system.
For this model, we consider all three qubits have the same transition frequency $\omega_0^i = \omega_0$.
The common environment is modelled as a set of bosonic field modes with frequencies $\omega_{k}$; the operators $b_{k}$ and $b_{k}^{\dagger}$ are the annihilation and creation operators associated with the $k^{th}$ mode of the environment.
In this model, we begin with a factorized initial state where the system and environment are decoupled, and the common
environment is in thermal equilibrium at some temperature $T$. For the tripartite system, the quantum master equation denotes
the decay dynamics under a common dephasing environment at a finite temperature governed by
\begin{eqnarray}
\frac{d}{dt} \rho(t) = -\frac{i}{\hbar} \big[ H_S, \rho(t) \big] +
\gamma(t) S_z \rho(t) S_z -\alpha(t) S_z S_z \rho(t) - \alpha^{\ast}(t) \rho(t) S_z S_z,
\label{Nc}
\end{eqnarray}
where $\gamma(t)$ is given by equation (\ref{gammat}) with an appropriate common spectral density $J(\omega)$ and
\begin{eqnarray}
\alpha(t) = \int_{0}^{\infty} d\omega  J(\omega) \coth\left(\frac{\hbar \omega}{2 k T}\right) \frac{\sin(\omega t)}{\omega}
- i \int_{0}^{\infty} d\omega  J(\omega) \frac{1-\cos(\omega t)}{\omega}.
\label{alphat}
\end{eqnarray}

\noindent In this work, we investigate the temperature dependence of coherence dynamics in environments 
characterized by an Ohmic spectral density $J(\omega)=\eta \omega \exp (-\omega/\Lambda)$. 
For numerical simulations, we vary the initial temperature of the thermal reservoir $kT$ in unit of $\hbar \omega_0$. 
The other parameter values are taken as $\eta=0.1\omega_0$, and $\Lambda=0.1\omega_0$. 
Throughout this work we consider the weak-coupling regime with $\eta=0.1\omega_0$,
consistent with previous investigations of spin-boson models \cite{goa2010,chenpwali2014,addis2014,aliphysicaa2024,alihome2026}.
The narrow cutoff frequency represents a typical structured reservoir employed in studies of finite temperature time-dependent
dephasing. Since the present work focuses on the influence of environmental temperature on multipartite coherence, these
representative parameter values are used throughout unless otherwise stated. Next, we investigate the coherence dynamics of
various tripartite initial states under both LDE and CDE. We emphasize that the dephasing rate $\gamma(t)$, given by
Eq.~(\ref{gammat}), as well as the time dependent  coefficient $\alpha(t)$ appearing in the above quantum master
equation, explicitly depend on thermal factors. This intrinsic temperature dependence establishes a direct connection
between coherence dynamics and the surrounding thermal environment. Consequently, the time-resolved
decay profile of quantum coherence provides the physical basis for coherence based quantum thermometry.
In multipartite systems, the interplay between state structure and \textit{system-environment} coupling leads to distinct temperature
sensitivities, resulting in state dependent modulation of thermal decoherence. This feature is directly relevant to quantum
calorimetry, where engineered quantum platforms probe microscopic heat exchange. Accordingly, analyzing coherence
under thermal variations not only clarifies decoherence mechanisms but also identifies regimes in which coherence itself
can serve as an effective temperature estimator in structured environments. \\

\section{States subjected to time-dependent dephasing noise}

\noindent Now we discuss the pure and mixed states that we have considered in this work for analysis.

\subsection{Pure states}

\vskip 0.5cm
\noindent We begin by considering the following representative three-qubit pure states, namely the
$\vert GHZ\rangle$, $\vert W\rangle$, $\vert W\overline{W}\rangle$, and $\vert Star\rangle$ states.
The structural characteristics of these states are as follows.
Tripartite pure states are classified into two inequivalent categories under Stochastic Local Operations and Classical
Communication (SLOCC), namely the $\vert GHZ \rangle$ class and the $\vert W \rangle$ class. Both the $\vert GHZ \rangle$
and $\vert W \rangle$ states exhibit genuine tripartite entanglement; however, they possess fundamentally different entanglement
structures. The $\vert GHZ \rangle$ state is highly fragile in the sense that tracing out any one qubit destroys all entanglement among the remaining qubits. In contrast, the $\vert W \rangle$ state exhibits a more robust entanglement distribution, whereby tracing out one
qubit still leaves the remaining two qubits entangled. These distinct structural characteristics lead to markedly different coherence
dynamics under local and common dephasing environments. The $\vert GHZ \rangle$ and $\vert W \rangle$ states are defined as \cite{kafatos2013bell,dur2000three}
\begin{eqnarray}
\vert GHZ\rangle &=& \frac{1}{\sqrt{2}}\Big(\vert 000\rangle + \vert 111\rangle\Big),\nonumber\\
\vert W\rangle &=& \frac{1}{\sqrt{3}}\Big(\vert 100\rangle + \vert 010\rangle + \vert 001\rangle\Big).
\label{ghzwpure_new}
\end{eqnarray}

\noindent In addition, we consider two other relevant tripartite states, namely the $\vert W\overline{W}\rangle$ and $\vert Star\rangle$ states \cite{roy2023exploring,cao2020fragility,royanu2025}. The $\vert W\overline{W} \rangle$ state is constructed as an equal coherent superposition of the $\vert W \rangle$ state and its spin-flipped counterpart $\vert \overline{W}\rangle$, given by
\begin{eqnarray}
\vert W\overline{W}\rangle = \frac{1}{\sqrt{2}}\Big(\vert W\rangle + \vert \overline{W}\rangle\Big),
\label{wwbarstate}
\end{eqnarray}
where
\begin{eqnarray}
\vert \overline{W}\rangle = \frac{1}{\sqrt{3}}\Big(\vert 011\rangle + \vert 101\rangle + \vert 110\rangle\Big).
\end{eqnarray}

The $\vert W\overline{W} \rangle$ state is particularly suitable for investigating coherence and correlations since it exhibits quantum coherence at single-qubit, bipartite, and tripartite levels, along with nontrivial bipartite and tripartite correlations. Thus, it serves as an effective platform for analyzing how quantum correlations are distributed across different subsystems. It is worth noting that the $\vert GHZ \rangle$, $\vert W \rangle$, and $\vert W\overline{W} \rangle$ states are symmetric under permutation of qubits. Therefore, the entanglement properties of their reduced bipartite states remain invariant regardless of which qubit experiences decoherence.\\

\noindent The $\vert Star \rangle$ state is somewhat different from the rest. This state is asymmetric in nature and displays an uneven
distribution of correlations. The $\vert Star \rangle$ state is defined as
\begin{eqnarray}
\vert Star\rangle = \frac{1}{2}\Big(\vert 000\rangle + \vert 100\rangle + \vert 101\rangle + \vert 111\rangle\Big).
\label{starstate}
\end{eqnarray}

\noindent Similar to the $\vert W\overline{W} \rangle$ state, the $\vert Star \rangle$ state also possesses coherence and
correlations at all hierarchical levels. However, due to its asymmetric structure, it consists of two distinct types of
qubits: (i) two peripheral qubits (the first and second) and (ii) one central qubit (the third). Consequently, the reduced
states obtained by tracing out different qubits exhibit distinct correlation properties. In particular, tracing out the central
qubit leaves the remaining two peripheral qubits in a separable state, highlighting the asymmetric distribution of quantum
coherence and correlations within the $\vert Star \rangle$ state. We examine these four tripartite pure states defined
above and investigate the evolution of their coherence under dephasing dynamics and subjected to finite temperature reservoir.

\subsection{Mixed states}

\noindent The three mixed states that we have taken into account here are
\begin{enumerate}
\item $\vert GHZ\rangle - \vert W\rangle$ mixture (denoted by $\rho^{W}_{GHZ}$),
\item Werner-$\vert GHZ \rangle$ state (denoted by $\rho^{GHZ}_{W\!E\!R}$),
\item Werner-$\vert W \rangle$ state (denoted by $\rho^{W}_{W\!E\!R}$),
\end{enumerate}
respectively. The mixture of the $\vert GHZ \rangle$ and $\vert W \rangle$ state, Werner-$\vert GHZ \rangle$ states,
and Werner-$\vert W \rangle$ states are defined as follows:
\begin{eqnarray}
\label{ghz-w}
\rho^{W}_{GHZ} = p\vert GHZ\rangle\langle GHZ\vert + (1-p)\vert W\rangle\langle W\vert,
\end{eqnarray}
\begin{eqnarray}
\label{werner-ghz}
\rho^{GHZ}_{W\!E\!R} = p\vert GHZ\rangle\langle GHZ\vert + \frac{1-p}{8}\mathbb{I}_{8},
\end{eqnarray}
and
\begin{eqnarray}
\label{werner-w}
\rho^{W}_{W\!E\!R} = p\vert W\rangle\langle W\vert + \frac{1-p}{8}\mathbb{I}_{8}.
\end{eqnarray}
Here, $p$ is the probability of mixing and $\mathbb{I}_{8}$ is the identity matrix of order $8$. The states are
defined in Eqs.(\ref{ghz-w}), (\ref{werner-ghz}) and (\ref{werner-w}). \\

\noindent The mixed states considered here constitute representative benchmark states that have been extensively
used in studies of multipartite entanglement, quantum coherence, and decoherence in open quantum systems.
Werner-type states provide an effective description of imperfect state preparation and depolarizing noise,
whereas mixtures involving the $\vert GHZ \rangle$ and $\vert W \rangle$ states facilitate systematic
investigations of the robustness of inequivalent multipartite entanglement structures under decoherence. Such
benchmark states \cite{dur2000three,roy2023exploring,royanu2025,werner1989quantum,acin2000generalized,
pan2003experimental,bourennane2004experimental,brockerhoff2026contrasting} therefore provide a physically
motivated framework for comparing the thermal susceptibility of multipartite coherence under various
environmental conditions.

\section{Results and Discussion}\label{sec:results}

\noindent In this section, we present a systematic analysis of the finite temperature coherence dynamics of representative multipartite pure and mixed states under local and common dephasing environments. By comparing different environmental configurations and state architectures, we investigate how thermal fluctuations influence coherence decay, stationary coherence, and long time dynamics. Particular emphasis is placed on identifying the physical mechanisms underlying the observed state dependent thermal response and on establishing the coherence temperature correspondence that forms the basis of coherence based quantum thermometry.

\begin{figure*}[h]
\includegraphics[width=16.5cm]{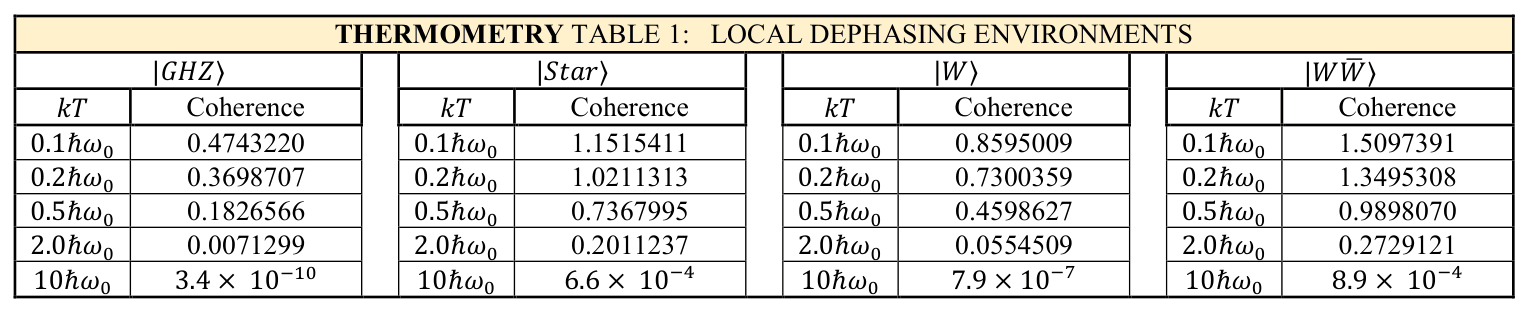}
\includegraphics[width=16.5cm]{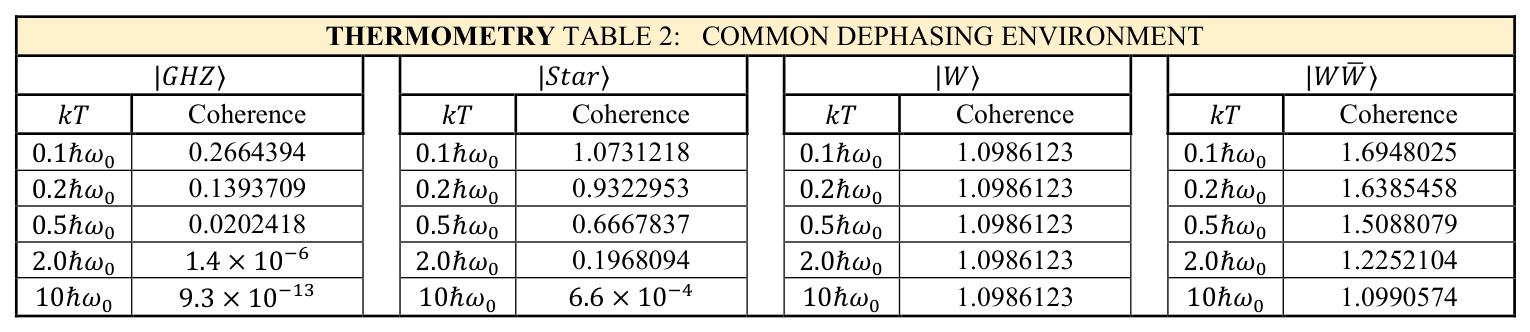}
\caption{Representative coherence thermometry tables for multipartite quantum systems evolving under (top) local and (bottom)
common dephasing environments. The relative entropy of coherence is evaluated at the fixed evolution time $\omega_0 t=3$
for the representative tripartite pure states $\vert GHZ\rangle$, $\vert Star\rangle$, $\vert W\rangle$, and $\vert W\overline{W}\rangle$.
The tabulated coherence values establish a direct correspondence between the measured quantum coherence and the reservoir temperature, thereby providing the calibration data for the \textit{proof-of-principle} coherence based thermometry discussed in the text.}
\label{fig2}
\end{figure*}

\noindent Before presenting the complete temperature dependence of multipartite coherence, we show here coherence thermometry of the pure states under local and common dephasing thermal environments. We already pointed out that the time dependent  dephasing rates appearing in the quantum master equation depend explicitly on the temperature of the surrounding reservoir through thermal factors. Consequently, the coherence dynamics of the multipartite quantum system become intrinsically linked to the environmental temperature. This direct dependence establishes quantum coherence as a temperature sensitive observable, whose measured value carries information about the thermal state of the reservoir. The primary objective of the present work is to investigate this coherence temperature relationship and examine whether multipartite quantum coherence can serve as an indicator of the surrounding thermal environment. To illustrate this relationship, we provide representative thermometry tables of Pure states (\textcolor{blue}{See Fig.~$2$}) for both local and common dephasing environments. The coherence is evaluated at a fixed evolution time, chosen here as $\omega_0 t=3$, for several representative multipartite pure states. For the parameter regime considered in this work, the measured coherence $C_R(\rho)$
at this observation time uniquely corresponds to a specific reservoir temperature $T$. Therefore, once the coherence is experimentally determined, the environmental temperature can, in principle, be inferred by interpolating or extrapolating the tabulated coherence temperature data. These tables thus serve as calibration tables establishing a direct correspondence between the relative entropy of coherence and the reservoir temperature.\\

\noindent The first Table (\textcolor{blue}{i.e. Table 1 of Fig.~$2$}) presents the coherence temperature correspondence for the $\vert GHZ\rangle$, $\vert Star\rangle$, $\vert W\rangle$, and $\vert W\overline{W}\rangle$ states evolving under local dephasing environments. A systematic decrease of coherence with increasing temperature is observed for all four states, demonstrating that thermal fluctuations progressively suppress multipartite quantum coherence. The rate of coherence degradation, however, depends strongly on the initial state, indicating that different multipartite state architectures possess markedly different thermal susceptibilities. This state dependent thermal response demonstrates that coherence can effectively encode information about the reservoir temperature. For comparison, the second Table (\textcolor{blue}{i.e. Table 2 of Fig.~$2$}) summarizes the corresponding coherence temperature relationship for the same states evolving under a common dephasing environment. In contrast to the local environment, collective coupling to the common reservoir produces qualitatively different thermal responses. While the $\vert GHZ\rangle$ and $\vert Star\rangle$ states exhibit pronounced temperature dependent coherence decay, the $\vert W\rangle$ state maintains a temperature independent stationary coherence owing to its symmetry under collective dephasing. The $\vert W\overline{W}\rangle$ state displays an intermediate behaviour, where the coherence decreases with increasing temperature but approaches a finite asymptotic value at long times. These observations demonstrate that the thermal susceptibility of multipartite coherence is governed not only by environmental temperature but also by the interplay between the internal architecture of the quantum state and the nature of the \textit{system-environment} interaction.\\

\noindent The representative thermometry tables shown in Fig.~$2$ are presented only for the pure multipartite states for illustration. An analogous coherence temperature correspondence can be established for the mixed states under both local and common dephasing environments. Consequently, similar thermometry tables can be constructed for the mixed states, providing the corresponding calibration data for coherence based temperature estimation. The thermometry tables therefore provide a direct operational connection between quantum coherence and environmental temperature for the present model. Rather than introducing a complete quantum thermometric protocol, the present work establishes the physical basis for coherence based thermometry by demonstrating a one-to-one correspondence between coherence and temperature in finite temperature environments. In this sense, quantum coherence serves as a measurable temperature sensitive observable that can, in principle, be employed to probe structured thermal environments.

\subsection{Pure States}

\noindent  We first analyze how the coherence of pure states defined in Eqs.~(\ref{ghzwpure_new}), (\ref{wwbarstate}) and (\ref{starstate}) behave under local dephasing environments.

\subsubsection{Pure States in Local Environment}

\begin{figure*}[h]
\includegraphics[width=\textwidth]{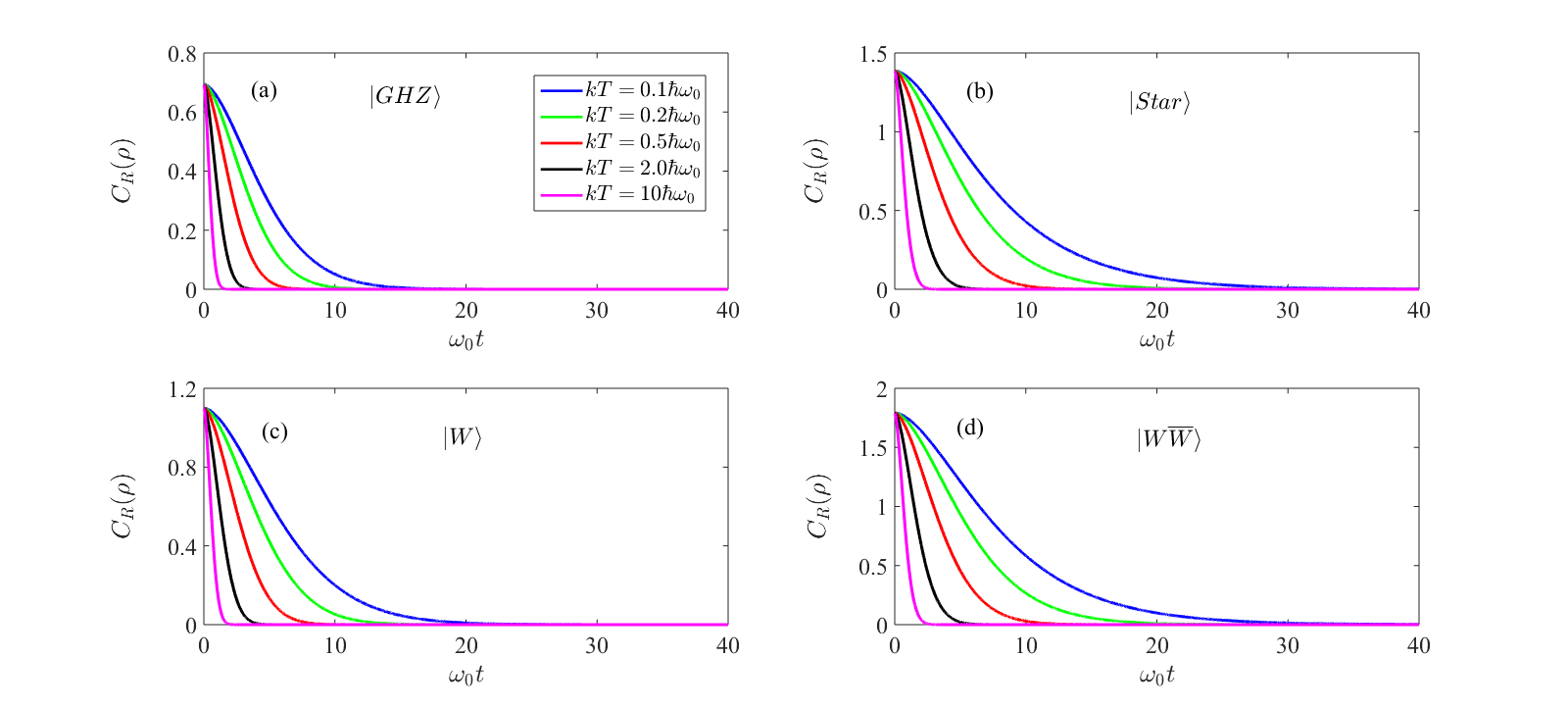}
\caption{\label{fig3} The dynamics of coherence of pure states under local dephasing environments. Here (a) represents the
relative entropy of coherence $C_R(\rho)$ for the initial state $\vert GHZ \rangle$ with varying environment
temperature $kT = 0.1\hbar\omega_0,~0.2\hbar\omega_0,~0.5\hbar\omega_0,~2\hbar\omega_0,~10\hbar\omega_0$;
(b) shows the dynamics of relative entropy of coherence $C_R(\rho)$ for the $\vert Star \rangle$-state under different
temperatures of the environment, while Figs.~(c) and (d) are showing dynamics of $C_R(\rho)$ for $\vert W \rangle$
state and $\vert W \overline{W} \rangle$ state respectively.}
\end{figure*}

\noindent We first investigate the influence of finite environmental temperature on the coherence dynamics of representative tripartite pure states evolving under local dephasing environments. The corresponding dynamics of the relative entropy of coherence $C_R(\rho)$ for the four initial states, namely the $\vert GHZ \rangle$, $\vert Star \rangle$, $\vert W \rangle$, and $\vert W \overline{W} \rangle$ states, are presented in
Fig.~\ref{fig3} for different values of the temperature $kT$.\\

\noindent A common feature observed in all four states is that the coherence decreases monotonically with time, and the rate of coherence decay increases systematically as the environmental temperature is raised. This behaviour originates from the explicit temperature dependence of the time dependent  dephasing rates appearing in the quantum master equation. Increasing the reservoir temperature enhances thermal fluctuations, which strengthen the phase randomization induced by the environment and accelerate the suppression of the off-diagonal elements of the density matrix. Consequently, the relative entropy of coherence decays more rapidly at higher temperatures. Thus, under local dephasing environments, finite temperature acts as a universal source of coherence degradation irrespective of the initial multipartite state.\\

\noindent Although the qualitative trend is common to all four states, the quantitative response to temperature depends strongly on the internal architecture of the multipartite state. Figure~\ref{fig3}(a) shows that the $\vert GHZ \rangle$ state exhibits the most pronounced thermal fragility. At the fixed observation time $\omega_0 t=3$, the coherence decreases from approximately $0.47$ at $kT=0.1\hbar\omega_0$ to $3.4\times 10^{-10}$ at $kT=10\hbar\omega_0$, indicating an almost complete suppression of coherence over the investigated temperature range. This strong temperature sensitivity originates from the fact that the coherence of the $\vert GHZ \rangle$ state is highly susceptible to independent phase fluctuations acting on individual qubits.\\

\noindent The $\vert Star \rangle$ state shown in Fig.~\ref{fig3}(b) displays a comparatively slower coherence decay than the $\vert GHZ \rangle$
state. Although thermal fluctuations continue to accelerate decoherence, the distributed coherence present in the $\vert Star \rangle$ state
results in improved robustness against local dephasing. Consequently, a larger fraction of the initial coherence survives over the same temperature range. The $\vert W \rangle$ state shown in Fig.~\ref{fig3}(c) exhibits an intermediate behaviour. As the temperature
increases, its coherence decreases systematically; however, the degradation remains less severe than that of the $\vert GHZ \rangle$ state.
The more delocalized distribution of quantum coherence among the three qubits leads to a reduced sensitivity to local thermal noise.\\

\noindent Among the four pure states considered, the $\vert W \overline{W} \rangle$ state shown in Fig.~\ref{fig3}(d) possesses the largest initial coherence and correspondingly maintains the largest coherence throughout the evolution. Even though the coherence decreases
monotonically with increasing temperature, the state retains appreciable coherence over a relatively broad temperature range, reflecting
its comparatively stronger resistance to thermal decoherence. This behaviour further demonstrates that the thermal susceptibility of
coherence depends not only on the environmental temperature but also on how quantum coherence is distributed within the multipartite state.\\

\noindent The results presented in Fig.~\ref{fig3} therefore establish two important physical observations. First, finite temperature universally accelerates decoherence under local dephasing environments through the temperature dependence of the dephasing rates. Second, the magnitude of thermal degradation is strongly state dependent, revealing that different multipartite state architectures possess markedly different thermal susceptibilities. This systematic and monotonic dependence of coherence on reservoir temperature establishes a direct coherence temperature correspondence and provides the physical basis for employing coherence as a temperature sensitive observable in the \textit{proof-of-principle} coherence based thermometry discussed in this work.

\subsubsection{Pure States in Common Environment}

\begin{figure*}[t]
\includegraphics[width=\textwidth]{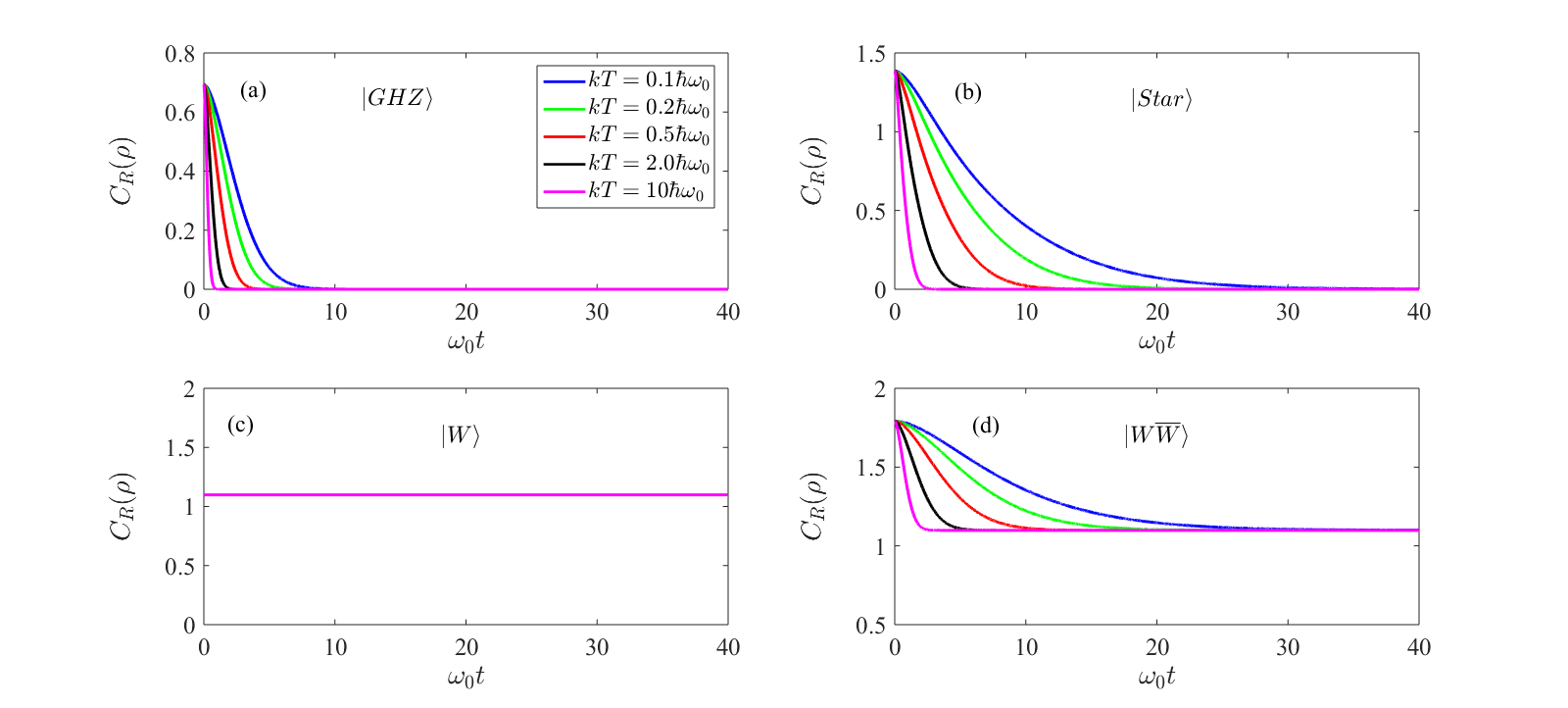}
\caption{\label{fig4} The dynamics of coherence of pure states under a common dephasing environment.
Here (a) represents the relative entropy of coherence $C_R(\rho)$ for initial state $\vert GHZ \rangle$ with varying
environment temperature $kT=0.1\hbar\omega_0,~0.2\hbar\omega_0,~0.5\hbar\omega_0,~2\hbar\omega_0,
~10\hbar\omega_0$; (b) shows the dynamics of relative entropy of coherence $C_R(\rho)$ for the $\vert Star \rangle$
-state under different temperatures of the environment, while Figs. (c) and (d) are showing dynamics of $C_R(\rho)$ for
$\vert W \rangle$ state and $\vert W \overline{W} \rangle$ state respectively.}
\end{figure*}

\noindent We next investigate the coherence dynamics of the same representative tripartite pure states when the three qubits interact collectively
with a common dephasing environment. The corresponding results are presented in Fig.~\ref{fig4} for different values
of the temperature $kT$. In contrast to the local dephasing environment, where all states exhibit qualitatively similar coherence decay,
the common environment produces markedly different dynamical behaviour depending on the symmetry and internal structure of the
multipartite quantum state.\\

\noindent The qualitative difference originates from the collective nature of the \textit{system-environment} interaction. Since
all three qubits couple
to the same reservoir, the environmental fluctuations become correlated, and the resulting coherence dynamics are governed not only
by the reservoir temperature but also by the symmetry of the multipartite state under the collective dephasing. Consequently,
states that are affected by local dephasing can display remarkably different thermal responses in the presence of a
common reservoir.\\

\noindent Figures~\ref{fig4}(a) and \ref{fig4}(b) show that both the $\vert GHZ \rangle$ and $\vert Star \rangle$ states undergo a monotonic
decay of coherence towards zero. Increasing the environmental temperature accelerates the coherence loss, indicating that thermal
fluctuations continue to enhance collective dephasing even in the presence of reservoir correlations. Compared with the local
environment, however, the coherence dynamics exhibit a different quantitative behaviour because the qubits no longer interact
with independent reservoirs but instead experience correlated phase fluctuations generated by the common environment.\\

\noindent A qualitatively different behaviour is observed for the $\vert W \rangle$ state, shown in Fig.~\ref{fig4}(c). Unlike the $\vert GHZ \rangle$
and $\vert Star \rangle$ states, the coherence remains essentially unchanged throughout the entire evolution and is practically independent of temperature. This remarkable robustness is a direct consequence of the symmetry of the $\vert W \rangle$ state under collective dephasing. The $\vert W \rangle$ state belongs to a decoherence-free subspace \cite{lidar1998decoherence,lidar1999concatenating} of the collective
dephasing interaction and is therefore protected against the phase fluctuations generated by the common reservoir. Consequently, although the master equation contains explicitly temperature-dependent dephasing rates, the corresponding coherence remains stationary because the collective dephasing leaves the $\vert W \rangle$ invariant. This result demonstrates that the thermal susceptibility of quantum coherence depends not only on environmental temperature but also on the symmetry properties of the multipartite quantum state.\\

\noindent Figure~\ref{fig4}(d) shows an intermediate behaviour for the $\vert W \overline{W} \rangle$ state. The coherence initially decreases due to environmental dephasing but subsequently approaches a finite stationary value that remains largely insensitive to further evolution. The asymptotic coherence is not a numerical coincidence but originates from the long-time steady state $\rho_{steady}$
$=1/2 \vert W \rangle \langle W \vert$ $+1/2 \vert \overline{W} \rangle \langle \overline{W} \vert$, whose relative entropy of coherence
is $C_R(\rho_{steady})=\ln 3 \approx 1.099$.\\

\noindent Thus, the observed stationary coherence follows directly from the structure of the asymptotic density matrix. Although increasing
temperature accelerates the initial coherence decay, the stationary value remains determined by the symmetry of the long-time state
rather than by thermal fluctuations. This behaviour illustrates that collective environmental coupling can preserve a finite amount
of quantum coherence even after long interaction times.\\

\noindent The results presented in Fig.~\ref{fig4} reveal a fundamentally different thermal response from that observed under local dephasing.
While local environments universally degrade coherence for all multipartite states, the common environment produces strongly
state dependent behaviour ranging from complete thermal decoherence to nearly temperature independent stationary coherence.
These observations demonstrate that the thermal susceptibility of multipartite coherence is jointly determined by the environmental
configuration and the internal architecture of the quantum state. Consequently, coherence measured under collective environmental
coupling contains information not only about the reservoir temperature but also about the symmetry of the underlying multipartite
state.

\subsection{Mixed states}

\noindent We next investigate the mixed states defined in Eqs.~(\ref{ghz-w}), (\ref{werner-ghz}), and (\ref{werner-w}), 
and analyze how their coherence evolves under time-dependent dephasing at finite temperature.

\subsubsection{Mixed States in Local Environment}

\begin{figure*}[t]
\centering
\includegraphics[width=\textwidth]{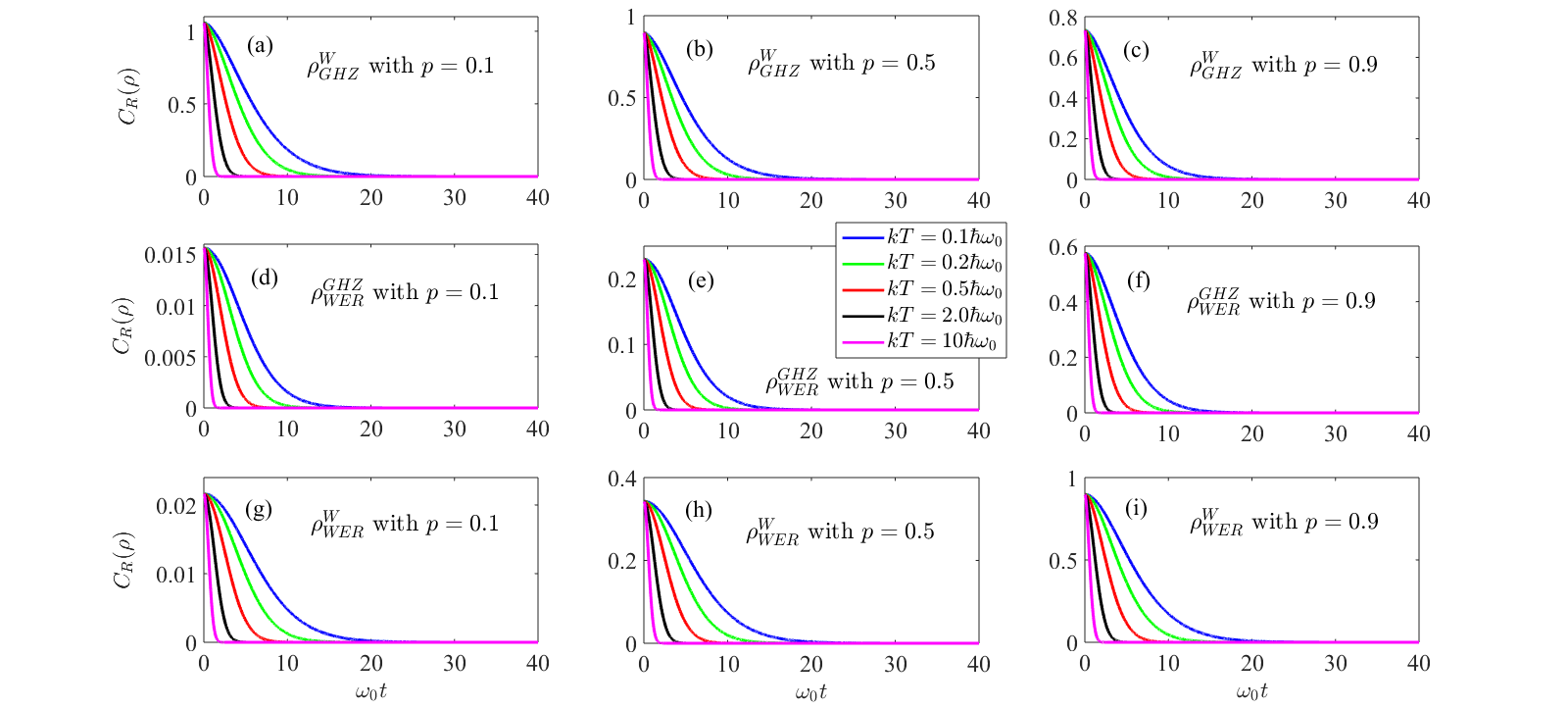}
\caption{\label{fig5} The dynamics of coherence of mixed states under local dephasing environments. Here (a)-(c) represent the relative entropy of coherence $C_R(\rho)$ of the mixed state $\rho_{GHZ}^{W}$ for mixing parameter $p=0.1$, $p=0.5$ and $p=0.9$ respectively, with varying environment temperature $kT = 0.1\hbar\omega_0,~0.2\hbar\omega_0,~0.5\hbar\omega_0,~2\hbar\omega_0,~10\hbar\omega_0$; Figs.~(d)-(f) show the relative entropy of coherence $C_R(\rho)$ for the mixed state $\rho_{W\!E\!R}^{GHZ}$ with the same mixing parameters and temperatures, while Figs.~(g)-(i) show the dynamics of $C_R(\rho)$ for the mixed state $\rho_{W\!E\!R}^{W}$ respectively.}
\end{figure*}

\noindent To examine whether the temperature-dependent coherence dynamics observed for pure states remain valid in more realistic situations,
we next investigate three representative mixed tripartite states evolving under local dephasing environments. The
corresponding coherence dynamics are shown in Fig.~\ref{fig5} for the mixed states $\rho^{W}_{GHZ}$, $\rho^{GHZ}_{W\!E\!R}$,
and $\rho^{W}_{W\!E\!R}$ with different values of the mixing parameter $p$. Studying mixed states is particularly relevant because
imperfections in state preparation and unavoidable environmental interactions often lead to mixed rather than perfectly pure
quantum states in practical experimental implementations. \\

\noindent As in the case of pure states, all mixed states exhibit a monotonic decrease of coherence with time, and the coherence decay becomes progressively faster as the environmental temperature increases. This behaviour again reflects the explicit temperature dependence
of the time dependent  dephasing rates appearing in the quantum master equation. Increasing thermal fluctuations enhance phase
randomization in the local reservoirs, leading to a more rapid suppression of the off-diagonal elements of the density matrix and,
consequently, a faster reduction of the relative entropy of coherence. These results demonstrate that finite temperature universally
accelerates decoherence under local dephasing irrespective of whether the initial state is pure or mixed.\\

\noindent The coherence dynamics of the three mixed states nevertheless exhibit significant quantitative differences, reflecting their distinct
internal structures. Figure~\ref{fig5}(a-c) shows the behaviour of the mixed state $\rho^{W}_{GHZ}$. Although the coherence
decreases monotonically for all values of the mixing parameter, the initial coherence depends sensitively on the relative
contributions of the $\vert GHZ \rangle$ and $\vert W \rangle$ components. As the temperature increases, the coherence
decreases systematically over the entire temperature range. For example, at the fixed observation time $\omega_0 t=3$, the coherence 
decreases from approximately $0.82$ at $kT=0.1\hbar\omega_0$ to $7.1 \times 10^{-7}$ at $kT=10\hbar\omega_0$ for $p=0.1$, 
 demonstrating the pronounced sensitivity of the coherence to temperature in this mixed state. \\

\noindent Figures~\ref{fig5}(d-f) and \ref{fig5}(g-i) present the corresponding results for the Werner-type mixed states $\rho^{GHZ}_{W\!E\!R}$
and $\rho^{W}_{W\!E\!R}$, respectively. In both cases, increasing the parameter $p$ increases the contribution of the coherent multipartite
component and therefore leads to larger initial coherence. Nevertheless, irrespective of the initial coherence, increasing the environmental temperature consistently accelerates the coherence decay. The quantitative differences among the three mixed states demonstrate that
thermal susceptibility depends not only on the reservoir temperature but also on the composition of the mixed state and the distribution
of coherence among its constituent components.\\

\noindent An important observation emerging from Fig.~\ref{fig5} is that the coherence temperature correspondence established for pure states
remains equally valid for mixed multipartite systems. Although the absolute magnitude of coherence depends on the degree of mixing,
the coherence decreases systematically with increasing temperature for all mixed states considered. Consequently, the measured
coherence continues to provide direct information about the surrounding thermal environment. This correspondence indicates that the
proposed coherence based thermometry is not restricted to idealized pure states but extends naturally to physically relevant mixed
states that are more likely to be encountered in realistic quantum experiments.\\

\noindent Overall, Fig.~\ref{fig5} demonstrates that finite environmental temperature acts as a universal mechanism for coherence degradation in
locally dephasing multipartite systems, while the magnitude of the thermal response remains strongly dependent on the internal
architecture and composition of the quantum state. These results further reinforce the central conclusion of the present work that
multipartite quantum coherence is a temperature sensitive observable whose thermal susceptibility is determined jointly by the
environmental conditions and the structure of the underlying quantum state.

\subsubsection{Mixed States in Common Environment}

\begin{figure*}[h]
\includegraphics[width=\textwidth]{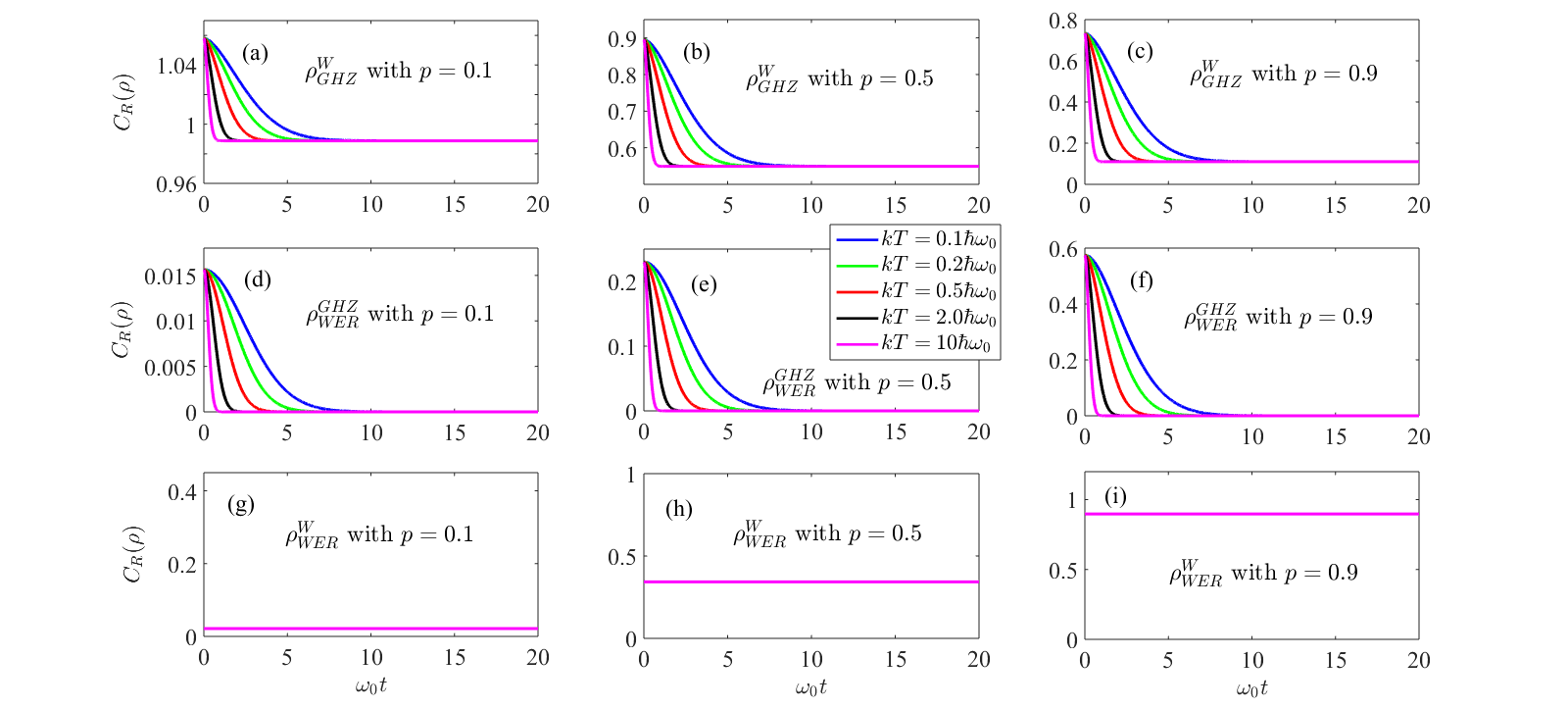}
\caption{\label{fig6} The dynamics of coherence of mixed states in a common dephasing environment. Here Figs. (a)-(c) represents the relative entropy of coherence $C_R(\rho)$ of the mixed state $\rho_{GHZ}^{W}$ for mixing parameter $p=0.1$, $p=0.5$ and $p=0.9$ respectively, with varying environment temperature $kT = 0.1\hbar\omega_0,~0.2\hbar\omega_0,~0.5\hbar\omega_0,~2\hbar\omega_0,~10\hbar\omega_0$; Figs. (d)-(f) show the relative entropy of coherence $C_R(\rho)$ for the mixed state $\rho_{W\!E\!R}^{GHZ}$ with the same mixing parameters and temperatures, while Figs. (g)-(i) are showing dynamics of $C_R(\rho)$ for mixed state $\rho_{W\!E\!R}^{W}$ respectively.}
\end{figure*}

\noindent Finally, we investigate the coherence dynamics of the mixed multipartite states when the three qubits interact collectively
with a common dephasing environment. The corresponding results are presented in Fig.~\ref{fig6} for the
three representative mixed states $\rho^{W}_{GHZ}$, $\rho^{GHZ}_{W\!E\!R}$, and $\rho^{W}_{W\!E\!R}$. In contrast to
the local dephasing environment, the collective interaction with a common reservoir gives rise to a much richer variety of
coherence dynamics, reflecting the combined influence of environmental correlations and the internal structure of the
mixed quantum states.\\

\noindent As in the pure-state case, the common environment introduces correlated phase fluctuations that affect all three qubits
simultaneously. Consequently, the coherence dynamics are no longer determined solely by the environmental temperature
but also by the symmetry properties and coherence distribution of the multipartite state. This interplay between collective
\textit{system-environment} coupling and state architecture produces markedly different thermal responses for different mixed
states, demonstrating that the influence of temperature cannot be understood independently of the underlying
quantum-state structure.\\

\noindent Figure~\ref{fig6}(a-c) present the coherence dynamics of the mixed state $\rho^{W}_{GHZ}$ under a common
dephasing environment. For $p=0.1$, $\vert W \rangle$ component to the initial state is more, the state is less sensitive 
to dephasing. The coherence usually decreases due to environmental dephasing and can also acquire a finite stationary 
value. The magnitude of this stationary coherence depends on the mixing parameter $p$, reflecting the relative contributions 
of the $\vert GHZ \rangle$ and
$\vert W \rangle$ components to the initial state. Increasing the reservoir temperature accelerates the initial coherence
decay; however, a finite amount of coherence survives even at long evolution times owing to the contribution from the
$\vert W \rangle$ component, which is protected against collective dephasing. This behaviour illustrates how the
composition of the mixed state directly determines its thermal robustness.\\

\noindent The behaviour becomes qualitatively different for the Werner-type mixed states. Figures~\ref{fig6}(d-f) show that the
coherence of $\rho^{GHZ}_{W\!E\!R}$ decays monotonically towards zero for all values of the mixing parameter.
Although increasing $p$ increases the initial coherence, finite temperature progressively suppresses the coherence until
complete decoherence is achieved at long times. This result indicates that the $\vert GHZ \rangle$ component remains
highly vulnerable to collective thermal dephasing despite the presence of environmental correlations.\\

\noindent In contrast, Figs.~\ref{fig6}(g-i) demonstrate that the mixed state $\rho^{W}_{W\!E\!R}$ exhibits robust coherence
throughout the evolution for all values of $p$. This behaviour originates from the fact that both the $\vert W \rangle$
state and the maximally mixed state $\mathbb{I}_{8}$ are invariant under the collective dephasing dynamics.
Since the Werner-$\vert W \rangle$ state is a convex mixture of these two invariant states, it also remains unaffected
by the common dephasing environment. Consequently, its coherence is essentially insensitive to both the interaction
time and the reservoir temperature.\\

\noindent The comparison among the three mixed states clearly demonstrates that the thermal susceptibility of multipartite
coherence under collective dephasing is strongly state dependent. States containing significant $\vert W \rangle$
-type contributions exhibit robust stationary coherence, whereas $\vert GHZ \rangle$-dominated states remain
highly susceptible to thermal decoherence. Thus, even though all states evolve under the same temperature
environment, their coherence dynamics differ substantially because of their distinct internal architectures and
symmetry properties. Under common dephasing, environmental temperature alone does not determine the
coherence dynamics; rather, the observed behaviour emerges from the interplay between thermal noise,
collective \textit{system-environment} coupling, and the internal structure of the multipartite quantum state.\\

\noindent The persistence of a clear coherence temperature relationship across both pure and mixed states demonstrates that
quantum coherence provides a robust temperature sensitive observable for structured thermal environments.
These observations further support the \textit{proof-of-principle} coherence based thermometry proposed in this work
and suggest that suitably engineered multipartite states may serve as useful probes of finite temperature
environments in future quantum thermometric and nanoscale calorimetric applications.

\section{Conclusion}

\noindent In this work, we have investigated the influence of finite environmental temperature on the coherence dynamics of multipartite quantum systems subjected to time-dependent dephasing. Using a tripartite spin-boson model, we analyzed the evolution of the relative entropy of coherence for several representative pure and mixed states under both local and common dephasing environments. The central objective of this study was to understand how environmental temperature is encoded in coherence dynamics and to examine whether quantum coherence can serve as a useful indicator of the thermal properties of the surrounding environment. Our results demonstrate that temperature plays a fundamental role in governing multipartite coherence dynamics. Under local dephasing environments, coherence exhibits a systematic and monotonic decrease with increasing temperature for all states considered, indicating a direct correspondence between thermal fluctuations and coherence degradation. In contrast, common dephasing environments generate markedly different behaviours depending on the structure of the multipartite state. While some states display strong temperature induced decoherence, others exhibit robust stationary coherence that remains largely insensitive to thermal fluctuations. These findings reveal that the thermal response of quantum coherence is highly state dependent and is determined by the interplay between the architecture of the quantum state and the nature of the \textit{system-environment} interaction. A principal outcome of this work is the identification of coherence as a temperature sensitive observable in finite temperature open quantum systems. Since the dephasing rates appearing in the quantum master equation depend explicitly on thermal factors, environmental temperature becomes directly encoded in the observed coherence dynamics. By fixing the observation time, we demonstrate a one to one correspondence between coherence and temperature for a variety of multipartite states, thereby providing a physical basis for coherence based thermometry. The thermometry tables presented in this work illustrate how environmental temperature may, in principle, be inferred from experimentally accessible coherence measurements. We emphasize that the present study is intended as a \textit{proof-of-principle} investigation rather than a complete quantum metrology framework. Our objective is not to establish ultimate thermometric precision or estimation bounds, but rather to demonstrate that coherence carries measurable and state dependent information about the thermal environment. In this sense, the work identifies multipartite coherence as a potentially useful resource for probing structured thermal environments and for exploring coherence based approaches to quantum thermometry and nanoscale quantum calorimetry. More broadly, our results highlight the importance of thermal effects in multipartite open quantum systems and demonstrate that different quantum state architectures can exhibit dramatically different thermal susceptibilities, ranging from rapid decoherence to temperature insensitive stationary coherence. These findings provide new insight into the interplay between quantum coherence, environmental temperature, and reservoir induced correlations. Future investigations incorporating quantum Fisher information, sensitivity analysis, and parameter estimation theory will be valuable for quantitatively assessing the metrological performance of multipartite coherence as a resource for temperature estimation.

\vskip 0.5cm
{\bf Declaration of competing interest} The authors declare that they have no known competing financial interests or
personal relationships that could have appeared to influence the work reported in this paper.

\vskip 0.5cm
{\bf Data availability statement} All data that support the findings of this study are included within the article. No supplementary file has been added.

\end{document}